\documentclass[useAMS,usenatbib]{mn2e}

\usepackage{graphicx,amssymb,rotating}
\usepackage{times}

\title[Photoionization rates around T Tauri Stars]{Constraints on the ionizing flux emitted by T Tauri stars}

\author[R.D.~Alexander, C.J.~Clarke \& J.E.~Pringle]
  {R.D.~Alexander\thanks{email: rda@ast.cam.ac.uk},
  C.J.~Clarke
  and J.E.~Pringle \\
   Institute of Astronomy, Madingley Road, Cambridge, CB3 0HA}

\begin{document}

\pagerange{\pageref{firstpage}--\pageref{lastpage}} \pubyear{2004}

\maketitle

\label{firstpage}

\begin{abstract}
We present the results of an analysis of ultraviolet observations of T Tauri Stars (TTS).  By analysing emission measures taken from the literature we derive rates of ionizing photons from the chromospheres of 5 classical TTS in the range $\sim 10^{41}$--$10^{44}$ photons s$^{-1}$, although these values are subject to large uncertainties.  We propose that the He\,{\sc ii}/C\,{\sc iv} line ratio can be used as a reddening-independent indicator of the hardness of the ultraviolet spectrum emitted by TTS.  By studying this line ratio in a much larger sample of objects we find evidence for an ionizing flux which does not decrease, and may even increase, as TTS evolve.  This implies that a significant fraction of the ionizing flux from TTS is not powered by the accretion of disc material onto the central object, and we discuss the significance of this result and its implications for models of disc evolution.  The presence of a significant ionizing flux in the later stages of circumstellar disc evolution provides an important new constraint on disc photoevaporation models.
\end{abstract}


\section{Introduction}
It is now well established that at an age of $\sim 10^6$ years, the majority of stars are surrounded by discs that are optically thick at optical and infrared wavelengths \citep*{strom89,kh95,haisch01}.  These discs are deduced from their sub-millimetre emission to be relatively massive (typically up to a few per cent of a solar mass; \citealt{beckwith90}) and are thus widely held to be potential sites of planet formation.  At an age of $\gtrsim 10^7$ years, however, most stars are no longer associated with such massive circumstellar discs, although low mass `debris discs' (believed to consist of dust replenished by cometary impacts) have been detected through submillimetre and scattered light imaging in some early type stars of this age \citep*{ms97,wyatt03}.  An outstanding unsolved question is: by what mechanism do stars lose their primordial discs as they approach the main sequence?  

Some clues about the nature of the disc dispersal mechanism are provided by the distribution of infrared colours among T Tauri stars \citep{kh95}.  These colours fall rather neatly into two categories: those that are explicable in terms of a stellar photosphere plus optically thick circumstellar disc and those that are compatible with purely photospheric emission.  (Note that these two categories tend to coincide with the spectroscopic designations of Classical and Weak Line T Tauri Star, henceforth CTTS and WTTS.)  As a number of authors have noted \citep*{skrutskie90,hartigan90,kh95,act99}, there is a striking lack of transition objects with colours intermediate between these two categories, suggesting that disc dispersal must occur on a time-scale that is {\it short} compared with the time the system spends as either a CTTS or a WTTS. 

Such two time-scale behaviour is however not readily produced by some of the most obvious mechanisms for disc dispersal, such as viscous evolution or magnetospheric clearing \citep{act99}.  These instead give rise to roughly power law declines in disc properties \citep{hcga98}, so that the time required to decline from any state is always of the same order as the time spent in the foregoing state.  Detailed models of disc photoevaporation were constructed by \cite{holl94,hollppiv} for the cases of both weak and strong stellar winds, but these models also fail to produce the required two-time-scale behaviour.  Subsequent studies have improved on the details of the photoevaporative winds, studying effects such as dust scattering \citep{ry97} and more detailed hydrodynamics \citep{font04}.  However the qualitative behaviour remains unchanged from that predicted by the initial models of \citet{holl94}.  Very recent work \citep{adams04} has shown that far-ultraviolet (FUV) photons, which can dissociate H$_2$ without ionizing neutral hydrogen, can also drive a significant disc wind.  However the study of \citet{adams04} focuses on an external radiation field, and it is not yet clear if a central FUV source can produce a similar effect.  The only model that has so far succeeded in achieving the required two time-scale behaviour combines viscous evolution with photoevaporation from the central star \citep*{cc01}.  In this case, although the star spends a relatively long time in the CTTS state (determined by the viscous timescale of the outer disc), at late times photoevaporation becomes important. This is effective only at radii beyond about 5 AU and the resulting mass loss from this region deprives the inner disc of re-supply.  Consequently, the inner disc drains on its own (short) viscous time-scale, and the time-scale for the disappearance of the inner disc is much less than the disc's lifetime to date.
   
However, as stressed by \citet{cc01} (see also \citealt*{mjh03,ruden04}), such a mechanism can only operate if the photoionizing radiation from the central star does not itself decline dramatically as the accretion onto the star is shut off.  This implies that there must be a significant contribution to the photoionizing radiation in T Tauri stars which is {\it not} powered by accretion if such photoevaporation models are to be successful.  TTS must not only produce a strong ionizing flux (of order $10^{41}$ ionizing photons per second), but must sustain this flux at the late stages of their evolution from CTTS to WTTS.

Currently, the origin of the photoionizing emission from TTS is unclear, and even its magnitude is poorly constrained \citep{gahm79,ia87}.  The rate of Lyman continuum photons emitted by the solar chromosphere is of order $10^{38}$s$^{-1}$ \citep{basri79,ayres97}, and studies of Herbig Ae/Be stars indicate ionizing fluxes of order $10^{43}$--$10^{45}$s$^{-1}$ \citep{bc98}, so it seems logical that the value for TTS lie between these two.  However no firm evidence exists regarding the ionizing continuum emitted by TTS.  One possible emission mechanism (see \citealt{cg98,lamzin98,gul00}) is that it is generated  mainly in an accretion shock where material from the disc impacts the stellar surface.  This ``accretion luminosity'' has been estimated to be capable of providing ionizing fluxes of order $10^{40}$--$10^{42}$ photon s$^{-1}$ \citep{hollppiv,mjh03,font04}, although this does not take into account the fact that the accretion column itself is highly optically thick to ionizing photons \citep*{columns}.  Further, it is evident that in this case the ionizing radiation would decline as the accretion rate declined, and photoevaporation would not be a viable mechanism for disc dispersal \citep{mjh03,ruden04}.  On the other hand, other authors have argued that this radiation instead derives from a scaled up version of solar-like magnetic activity \citep*{lago84,costa00}.  In this case, a decline in accretion rate would not be expected to shut off the source of photoionizing radiation, and photoevaporation models for disc dispersal would  in principle be viable.  To date, attempts to distinguish between these models on the basis of archival {\it IUE} spectra \citep*{costa00,jkvl00} have come to rather different conclusions about the relative contribution of accretion to the ultraviolet output of T Tauri stars.  In the X-ray domain, by contrast, it is evident that the main energy source is {\it not} accretion, since WTTS are, if anything, even more luminous in the X-rays than their CTTS counterparts \citep{damiani95,sn01}.  However, as long as the central object provides a strong UV radiation field X-rays are unlikely to be able to influence disc evolution significantly, as the mass-loss rates due to X-ray photoevaporation are much smaller than those driven by the fiducial UV photoevaporation rate \citep*{xrays}.

In this paper, we use archival data from {\it HST STIS} and from the {\it IUE} final archive \citep*{vjkl00} to re-examine the issue of the magnitude and origin of the photoionizing emission in T Tauri stars. The structure of the paper is as follows.  In Section \ref{sec:problems} we discuss the problems involved in estimating ionizing fluxes from observations, taking particular note of the large uncertainties which arise (primarily due to uncertainties in reddening).  In Section \ref{sec:EMs} we use an emission measure method to estimate the ionizing fluxes from a small sample of CTTS.  Following on from this (Section \ref{sec:idea}), we propose a simple, reddening independent, method of estimating the fraction of the UV power radiated at wavelengths short of the Lyman break, namely through the ratio of the He\,{\sc ii} 1640\AA~line to the C\,{\sc iv} 1550\AA~line.  Using data from {\it HST STIS} and from the {\it IUE} final archive \citep*{vjkl00} we examine how this ratio behaves as TTS evolve.  We find evidence that the relative contribution of high frequency radiation increases, or at least does not decrease, as the accretion rate on to the star declines and interpret this as evidence for an important chromospheric contribution to the ionizing flux in T Tauri stars.  In Section \ref{sec:dis} we discuss further caveats that apply to our technique, as well as some issues regarding source geometry, and summarise our conclusions in Section \ref{sec:summary}.

\section{Observational Problems}\label{sec:problems}
There are a number of problems involved in trying to determine the nature of the ionizing flux from T Tauri stars.  Direct observations of their spectra at wavelengths shortward of the Lyman break at 912\AA~are not possible as any escaping photons are readily absorbed by (abundant) interstellar neutral hydrogen.  Observations at slightly longer wavelengths ($\simeq1000$--2000\AA) tend to focus on line emission, and consequently have low continuum signal-to-noise.  This, combined with the large observational uncertainties in derived stellar parameters (such as distance and stellar radius; see, for example, \citealt{kh95}) and the uncertainty regarding what is ``true'' continuum emission and what is unresolved line emission, mean that the observed continua are not useful in constraining this problem further.  It is also impossible to evaluate the ionizing flux by studying the Balmer lines of neutral hydrogen, which can be due to radiative recombination, as previous studies \citep*{vbj93} have found that these lines are in local thermal equilibrium or, at the very least, that the excitation is collisionally dominated.

Correcting for reddening is also a serious issue in the ultraviolet (UV), as extinction is far greater here than in the optical or infra-red.  Previous observations have resulted in a range of reddening parameters being derived for these objects, with measurements of $A_V$ by different methods commonly differing by a magnitude or more.  (For T Tau observed values range from $A_V=0.8$ \citep{kl02} to $A_V=1.7$ \citep{gul00}.)  When extended to the UV such uncertainties, combined with uncertainties regarding the behaviour of the reddening law, result in very large uncertainties in the true UV fluxes of T Tauri stars.  We attempted to obtain independent estimates of the reddening towards our {\it HST} sources but this proved impossible by conventional methods.  Interstellar absorption lines such as the S\,{\sc ii} triplet at 1250\AA~are too weak and poorly resolved, and the rapid variation, on time-scales of weeks to days, of emission from T Tauri stars means that archival optical/UV data cannot be used due to the non-simultaneity of the observations.


\section{Emission Measure Analysis}\label{sec:EMs}
In order to make some quantitative estimates of the ionizing fluxes emitted by TTS we have first made use of an emission measure (EM) calculation.  In plasma physics, the differential emission measure (DEM) is defined as:
\begin{equation}\label{eq:dem}
DEM(T)=n_e n_H \frac{dh}{dT}
\end{equation}
where $n_e$ and $n_H$ are the particle number densities of electrons and hydrogen respectively, $T$ is temperature and $h$ is length along the line-of-sight.  Thus the intensity $I$ of a single spectral line due to a transition between atomic states $i$ and $j$ can be evaluated from the DEM as:
\begin{equation}
I\left(\lambda_{ij}\right) = Ab(X) \int_T C\left(T,\lambda_{ij},n_e\right) DEM(T) dT
\end{equation}
where $\lambda_{ij}$ is the wavelength of the line, $Ab(X)$ is the abundance of the element ($X$) in question and $C$ is the contribution function, which incorporates all the necessary atomic physics.  The DEM gives an indication of the amount of plasma emitting along the line-of-sight in the temperature interval $[T, T+dT]$, and the DEM and the necessary atomic data can therefore give intensities for a complete spectrum.  (For a detailed review of the necessary physics see \citealt{mmf94}.)  DEMs are derived from observed spectral data but the process is complex and difficult, requiring high-quality data.  Consequently few DEMs exist for TTS, as the spectral data observed are not typically of high enough quality to enable a DEM to be derived.

We have made use of the EMs for 5 CTTS derived by \citet{brooks01} (kindly provided in electronic form by David Brooks): BP Tau, RY Tau, RU Lup, GW Ori and CV Cha.  The analysis of \citet{brooks01} used line fluxes, observed with the {\it IUE} satellite, to calculate the distribution of the EM.  Each spectral line results in an ``emission measure locus'' centred around the formation temperature of the line, and so by sampling a series of lines over a range of formation temperatures \citet{brooks01} were able to produce an EM distribution as a function of temperature for each object studied.  The lines used were C\,{\sc ii} 1335\AA, the Si\,{\sc iv} doublet at 1393\AA, the C\,{\sc iv} doublet at 1549\AA, O\,{\sc iii]} 1660\AA, N\,{\sc iii]} 1752\AA, Si\,{\sc ii} 1816\AA, Si\,{\sc iii]} 1892\AA~and C\,{\sc iii]}1909\AA.  In addition the blended Si\,{\sc iii}/O\,{\sc i} pair at 1303\AA~and the N\,{\sc v} 1242\AA~line, which is blended with Ly$\alpha$, were also used.  This results in derived EM distributions that span the range $\log T\simeq 3.9$--5.6.

The EM distributions evaluated by \citet{brooks01} are volume EMs, and so in order to convert them to column EMs we have assumed spherical symmetry and scaled the distributions by a constant factor $4\pi R_*^2/d_*^2$ accordingly\footnote{A volume EM takes the same form as the column EM defined in equation \ref{eq:dem}, but with the length element $dh$ replaced by a volume element $dV$.}.  Here $R$ is the stellar radius and $d$ the Earth-star distance, and we have adopted the stellar parameters from \citet{bc03}.  We performed a simple polynomial fit to the EM loci to obtain single-valued DEMs as functions of temperature.  We then used the resulting scaled DEMs as inputs to the {\sc chianti} spectral synthesis code (version 4.01, \citealt*{chianti,chianti4}).  This code makes use of a large atomic database to simulate the thermal emission from optically thin plasmas.  It is useful in our study as it requires few input parameters, the most important of which is the EM distribution.  In this manner, we are able to generate synthetic spectra over a much larger wavelength range that can be observed.

\begin{figure}
        \resizebox{\hsize}{!}{
        \begin{turn}{270}
        \includegraphics{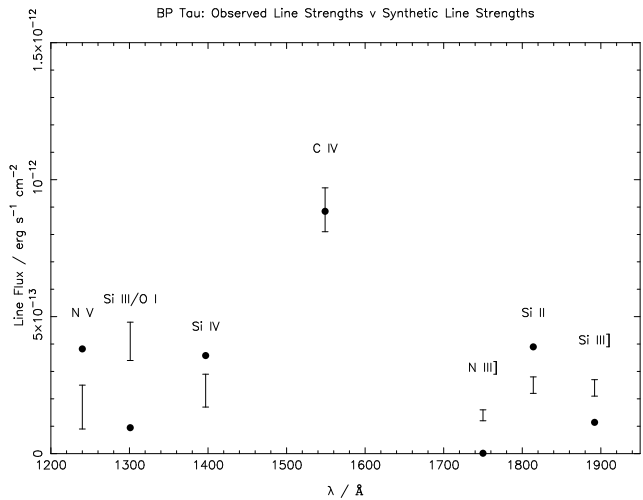}
        \end{turn}
        }
        \caption{Comparison between observed line strengths and the line strengths obtained from the EM model for BP Tau.  Observed strengths (the extinction-corrected values from \citealt{brooks01}) are shown as error bounds, with the synthetic values shown by filled circles.}
        \label{fig:bptau}
\end{figure}

As a test of the reliability of the procedure we have first attempted to reproduce the observed spectra from the models.  In order to achieve this we must correct the synthetic line fluxes for geometric dilution, again using the stellar parameters from \cite{bc03}.  We have focused on the strong, unblended lines, and found that in general we are able to reproduce the observed line strengths (the extinction corrected values listed in \citealt{brooks01}) to within a factor of 2--3.  Both the line ratios and the absolute strengths are in general well-matched.  (The exceptions to this are some of the forbidden lines which, as noted by \citet{brooks01}, are extremely density-sensitive and not always well matched to the observations.)  A comparison of the observed and calculated line strengths for a typical case (that of BP Tau) is shown in Fig.\ref{fig:bptau}.  

We note that the EM distributions are not always well-represented by a low-order polynomial, and in some cases the resulting line-fluxes are very sensitive to the fit.  The resultant spectrum also requires that the elemental abundances be specified.  Again, the line-fluxes are somewhat sensitive to the choice of abundances, but we are able to reproduce the observed values reasonably well using standard abundance data (the ``solar hybrid'' abundances of \citealt{fs99}).  We also note that the predicted strength of the C\,{\sc iv} 1550\AA~doublet is generally slightly too large, which may be an indication of opacity.  Lastly we note that the synthetic spectra completely fail to reproduce the He\,{\sc ii} 1640\AA~line, with predicted line strengths 2--3 orders of magnitude smaller than those observed.  (This line is not included in Fig.\ref{fig:bptau} as it was not used by \citet{brooks01} in the derivation of the EM and so no extinction-corrected value is quoted.)  This occurs despite the fact that the excitation temperature for this line is very close to that of the well-reproduced C\,{\sc iv} 1550\AA~line.  We attribute this to the fact that the observed He\,{\sc ii} 1640\AA~line emission is due to radiative recombination, rather than collisional excitation, and will return to this point later.

Whilst the spectra are generally well reproduced this is not true for GW Ori, where the predicted absolute line strengths are too large by a factor of $\simeq100$.  The error is approximately constant in all of the line strengths and so it appears to be a systematic error, probably due to the scaling of the EM.  The quoted radius for GW Ori (8.4R$_{\odot}$) is somewhat uncertain, but this uncertainty is unlikely to be responsible for such a large error.  We note that \citet{bc03} also found GW Ori to be anomalous, attributing their discrepancy to the failure of the assumption of spherical symmetry.  In addition we note that GW Ori is a spectroscopic binary \citep{mal91}, which may introduce further uncertainty.  As a result we have scaled down the EM for GW Ori by a further constant factor of 100 in order to reproduce the observed spectrum better.

As a further check we have attempted to reproduce the values of the ``total radiated power'' derived by \citet{bc03}.  Again, we can reproduce all these values to within the uncertainties introduced by the unknown parameters discussed above.  Thus we are satisfied that the EM distributions and the {\sc chianti} code provide a consistent way of synthesizing the spectra of the chromospheres of these TTS.

\begin{table}
 \centering
 \begin{minipage}{\hsize}
  \caption{Values of the photon fluxes derived for the 5 CTTS studied by \citet{brooks01} and \citet{bc03}.  The fluxes are presented in 3 wavebands: ionizing (700--912\AA), ``H$_2$-dissociating'' (912--1100\AA) and ``broadband FUV'' (912--2000\AA).  These values were derived by adopting the ionization equilibria from \citet{ar85} and \citet{ar92}, and ``solar hybrid'' elemental abundances \citep{fs99}; different choices can alter these values somewhat.  The extinction parameters $A_V$ adopted in deriving the emission measures are included for reference (from \citealt{bc03}).}\label{tab:phi}
  \begin{tabular}{lcccc}
  \hline
  Star     & \multicolumn{3}{c}{Photon flux in waveband $\left(\times10^{42}{\mathrm s}^{-1} \right)$} & $A_V$ \\
 & 700--912\AA~($\Phi$) & 912--1100\AA & 912--2000\AA & \\
 \hline
BP Tau  & 0.68 & 1.5 & 6.6 & 0.5 \\ 
RY Tau  & 2.1 & 2.3 & 14 & 0.55 \\
RU Lup  & 2.3 & 2.6 & 16 & 0.4 \\
GW Ori  & 13 & 9.4 & 56 & 0.8 \\
CV Cha  & 78 & 71 & 440 & 1.7 \\
\hline
\end{tabular}
\end{minipage}
\end{table}

In this manner we have used the {\sc chianti} spectral synthesis code to generate synthetic spectra of TTS chromospheres for the 5 sources mentioned above.  As a result, we can use these spectra to make estimates of the ionizing fluxes they produce.  We estimate this by summing the contributions to the synthetic spectra over the wavelength range 700--912\AA.  Extending the lower limit of this range to smaller values makes little difference to the total, as the Lyman continuum is dominated by photons at wavelengths very close to the Lyman break.  For comparison we also evaluate the photon fluxes in two wavebands longward of the Lyman break, 912--1100\AA~and 912-2000\AA.  The derived values are comparable to previous estimates of the FUV flux from TTS (eg.~\citealt{herczeg04}).  These estimates are subject to the modelling uncertainties mentioned above, and more significantly are also subject to the much larger uncertainties due to reddening.  As a result we consider these values to be order-of-magnitude estimates only.  However as shown in Table \ref{tab:phi}, the resulting ionizing fluxes are high enough to meet the demands of disc photoionization models, and in fact may be somewhat higher than the previously assumed fiducial value of $10^{41}$photon s$^{-1}$.  Thus we find that the Lyman continuum produced by TTS may indeed be large enough to drive disc photoionization models.  However a large ionizing flux alone is not sufficient; we also require that the ionizing flux does not decrease as the objects evolve.  Thus we now seek a diagnostic that will tell us about the evolution of the ionizing flux.


\section{A New Technique}\label{sec:idea}
As noted in Section \ref{sec:EMs}, whilst the synthetic spectra generated by the EM analysis are generally accurate, they fail completely to match the observed fluxes of the He\,{\sc ii} 1640\AA~line.  This line is the He\,{\sc ii} equivalent of H$\alpha$, and so may be produced by radiative recombination\footnote{The wavelength of the ``Lyman break'' for He\,{\sc ii} is 228\AA.}.  This would explain its absence from our synthetic spectra, which incorporate only collisional excitation of atoms and ions.

Bearing this in mind we propose a new technique for determining the behaviour of this ionizing flux.  If the He\,{\sc ii} 1640\AA~line is due to radiative recombination (there is some debate about this - see Section \ref{sec:heii_dis}), then the flux in the He\,{\sc ii} line should provide an indication as to the power of the ionizing radiation at wavelengths $<$ 228\AA.  In addition, \citet{bc03} found that the line flux in C\,{\sc iv} doublet at 1550\AA~is strongly correlated with the total power radiated by TTS atmospheres.  As a result, we propose that the He\,{\sc ii}/C\,{\sc iv} line ratio should provide an independent indicator of the relative flux of photons which are ionizing material around the stars in our sample.  We have investigated correlations between the observed line ratios (from \citealt{vjkl00}) and the ratio of He\,{\sc ii}-ionizing power to total power ratios evaluated from the EMs, for the 5 objects studied above.  Unfortunately the power ratio derived from the EM analysis is very sensitive to the slope of the EM distribution in the range $\log T \simeq4.2$--4.8.  The data from \citet{brooks01} show some scatter in this region, and as a result some of the derived power ratios are very sensitive to the fit.  Only BP Tau, RU Lup and CV Cha give robust power ratios for various fits.  The other 2 objects (RY Tau and GW Ori) produce power ratios that very uncertain, making the numbers useless in this regard.  As seen in Table \ref{tab:power_rat} the values for BP Tau, RU Lup and CV Cha do indicate that a harder spectrum produces an increased line ratio, supporting our hypothesis.

\begin{table}
 \centering
 \begin{minipage}{\hsize}
  \caption{Observed He\,{\sc ii}/C\,{\sc iv} line ratios and derived He\,{\sc ii}-ionizing/total power ratios for the 3 objects where the power ratio is robust.  Observed line ratios are taken from \citet{vjkl00}.}\label{tab:power_rat}
  \begin{tabular}{lcc}
  \hline
  Star     &  He\,{\sc ii}/C\,{\sc iv} &  Power at $\lambda < 228$\AA~/ Total power \\
 \hline
CV Cha & 0.12$\pm$0.04 & 0.0064 \\
RU Lup & 0.15$\pm$0.01 & 0.0054 \\
BP Tau & 0.50$\pm$0.01 & 0.0277 \\
\hline
\end{tabular}
\end{minipage}
\end{table}

The advantages of using this line ratio are twofold.  Firstly, it provides a ``normalised'' measurement, essentially measuring ``ionizing flux/total chromospheric power'' and so should be robust against variations between sources (which is important, as our sample covers a broad range of spectral types and evolutionary states).  Secondly, the fact that these lines are closely spaced in wavelength, separated by only 90\AA, means that the ratio is relatively insensitive to variations in reddening.  Indeed we found that adopting extreme values of both $A_V$ and $R_V$ (ie. the highest and lowest found in the literature) and dereddening observed spectra using the extinction curves of \citet*{ccm89} caused variations of less than 10\% in the line ratio: whilst the absolute line fluxes are very sensitive to extinction this line ratio is not.  In essence we propose that this line ratio represents a ``spectral hardness ratio'', related to the fraction of the total chromospheric power radiated at short wavelengths.  As can be seen from X-ray observations, there is no evidence that magnetic activity in TTS is related to disc evolution, as both CTTS and WTTS show similar levels of activity \citep{fm99}.  Thus studying the dependence of this line ratio against various evolutionary indicators should give an indication as to how the ionizing flux evolves.

\subsection{Observations and Data Reduction}
UV spectra of 9 CTTS were obtained from the Space Telescope Imaging Spectrograph ({\it STIS}) on board the Hubble Space Telescope ({\it HST}), taken from the {\it HST} public archive (see Table \ref{tab:stisdata} for the names of these objects).  Initial processing (flatfielding, cosmic ray removal etc.) was performed by the {\it HST} {\it STIS} reduction pipeline \citep{tr97} and so we were provided with flux-calibrated 2-dimensional spectra.  These spectra were added across the spatial extent of the targets and a sky subtraction was performed using an offset of 2$\arcmin$.  In cases where more than one spectrum was observed (see Table \ref{tab:stisdata}), the spectra were co-added to improve the S/N ratio.  The fluxes for the He\,{\sc ii} and C\,{\sc iv} lines were then extracted using the line-fitting routines from the {\sc figaro} package.  The He\,{\sc ii} 1640\AA~was fitted using a single Gaussian profile, whereas the C\,{\sc iv} line was fitted by 2 separate Gaussian components corresponding to the 1548\AA~and 1551\AA~peaks.  The results of these line fits are also shown in Table \ref{tab:stisdata}.

\begin{table}
 \centering
 \begin{minipage}{\hsize}
  \caption{Line strengths obtained from {\it HST} {\it STIS}.  Values for C\,{\sc iv} 1550\AA~are the sum of the two fitted components.}\label{tab:stisdata}
  \begin{tabular}{lcccc}
  \hline
  Star     &   No.~of       & 
\multicolumn{2}{c}{Flux ($\times 10^{-14}$ergs$^{-1}$cm$^{-2})$}  & He\,{\sc ii}/C\,{\sc iv}\\
 & Spectra & C\,{\sc iv} 1550 & He\,{\sc ii} 1640 & Ratio \\
 \hline
T Tau & 10 &  19.94$\pm$1.34 & 7.01$\pm$0.15 & 0.35$\pm$0.03 \\
RY Tau & 1 &  0.73$\pm$0.24 & 0.17$\pm$0.04 & 0.23$\pm$0.10 \\
SU Aur & 2 &  3.60$\pm$0.15 & 0.76$\pm$0.03 & 0.21$\pm$0.02 \\
GW Ori & 2 &  2.58$\pm$0.52 & 0.67$\pm$0.02 & 0.26$\pm$0.05 \\
CO Ori & 2 &  0.09$\pm$0.02 & 0.03$\pm$0.01 & 0.31$\pm$0.08 \\
EZ Ori & 1 &  0.75$\pm$0.05 & 0.14$\pm$0.01 & 0.19$\pm$0.02 \\
V1044 Ori & 1 &  1.58$\pm$0.09 & 0.26$\pm$0.02 & 0.16$\pm$0.02 \\
P2441 & 1 &  0.37$\pm$0.02 & 0.09$\pm$0.02 & 0.23$\pm$0.06 \\
RY Lup & 2 &  11.80$\pm$0.97 & 1.59$\pm$0.13 & 0.14$\pm$0.02 \\
\hline
\end{tabular}
\end{minipage}
\end{table}

\begin{figure}
        \resizebox{\hsize}{!}{
        \begin{turn}{270}
        \includegraphics{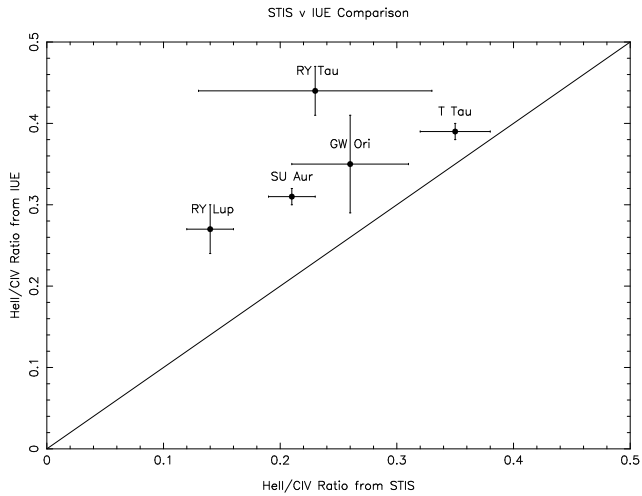}
        \end{turn}
        }
        \caption{Line ratios from {\it STIS} plotted against those from {\it IUE}, for the 5 objects observed by both satellites.  The solid line represents equality.}
        \label{fig:comp}
\end{figure}

The sample of 9 objects observed by {\it STIS} is small, and also covers a small spread in evolutionary properties, as all of the 9 objects are CTTS and have similar ages.  Further, the objects cover a broad range of spectral types (F8-K4) and so this sample is far from ideal.  In order to expand the sample we turned to data from the International Ultraviolet Explorer ({\it IUE}) final archive \citep{vjkl00}, which observed a total of 50 TTS, both classical and weak-lined.  In some cases the signal-to-noise ratio is very poor, so only 33 of these objects have reliable line fluxes for both C\,{\sc iv} 1550\AA~and He\,{\sc ii} 1640\AA.  The lower spectral resolution of {\it IUE} (compared to that of {\it STIS}) introduces some problems with line-blending especially, as noted by \citet{vjkl00}, in the He\,{\sc ii} 1640\AA~line.  However, plotting the He\,{\sc ii}/C\,{\sc iv} line ratio from {\it STIS} versus that from {\it IUE}, as shown in Fig.\ref{fig:comp}, seems to indicate that the line blends provide a systematic offset to this ratio.  If the variations were due to variability we would not expect to see such a systematic effect, and so it seems that whilst the absolute values of the line ratio differ between the two samples the appearance of any evolutionary trends should be not be affected by this systematic effect.  Consequently we use the ratios obtained from the {\it IUE} archive throughout the results section, as they are as valid as those from {\it STIS} and incorporate a much larger sample.

\subsection{Results}
\begin{figure}
        \resizebox{\hsize}{!}{
        \begin{turn}{270}
        \includegraphics{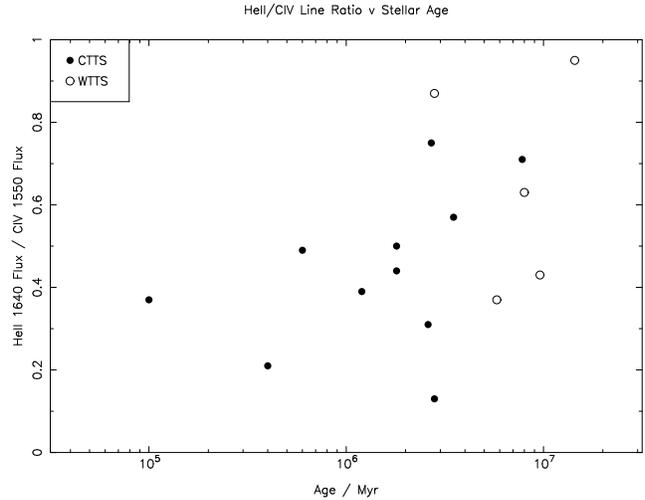}
        \end{turn}
        }
        \caption{Plot of line ratio versus stellar age. (Ages taken from \citealt{ps02}.)}
        \label{fig:ages}
\end{figure}

\begin{figure}
        \resizebox{\hsize}{!}{
        \begin{turn}{270}
        \includegraphics{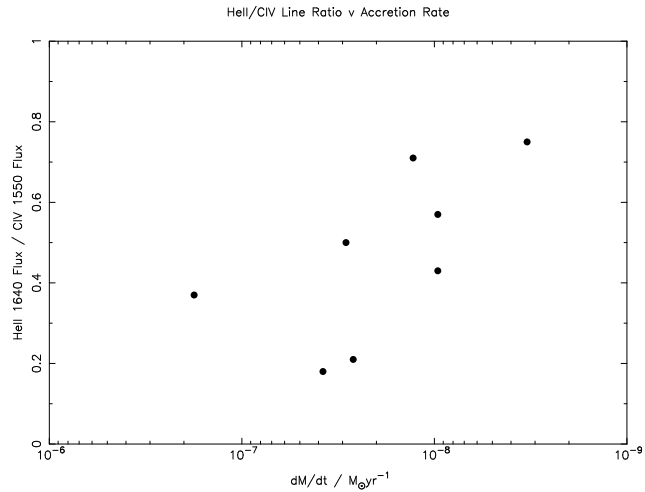}
        \end{turn}
        }
        \caption{Plot of line ratio versus accretion rate: objects ``evolve'' to the right. ($\dot{M}$ values from \citealt{gul98})}
        \label{fig:m_gul}
\end{figure}

\begin{figure}
        \resizebox{\hsize}{!}{
        \begin{turn}{270}
        \includegraphics{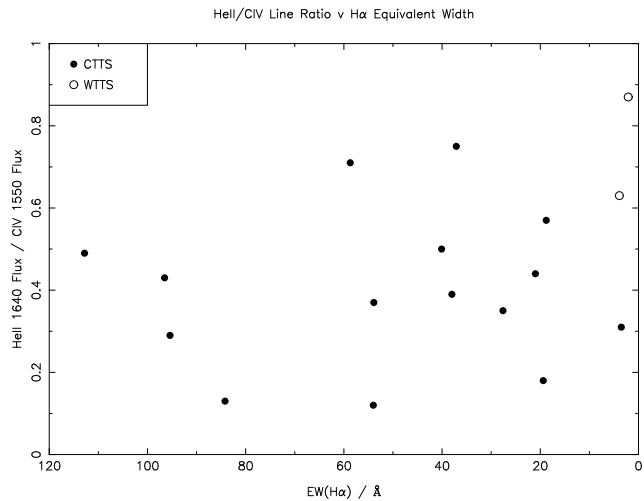}
        \end{turn}
        }
        \caption{Plot of line ratio versus H$\alpha$ equivalent width: objects ``evolve'' to the right. (Equivalent widths taken from \citealt{ck79}.)  SU Aur is a somewhat unusual object \citep{unruh04}, but it does posess a disc and so we have classified it as a CTTS despite its weak H$\alpha$ equivalent width of 3\AA.}
        \label{fig:ha}
\end{figure}

Figures \ref{fig:ages}-\ref{fig:ha} show the He\,{\sc ii}/C\,{\sc iv} line ratio (from the data in \citealt{vjkl00}) plotted against several evolutionary indicators (where available).  While good values of the line ratio are available for 33 objects, in some cases the evolutionary indicators are less well-known.  Thus not all of the objects appear on all of the plots.

Fig.~\ref{fig:ages} shows the line ratio plotted against stellar age, with values for the ages, derived from pre-main-sequence evolutionary tracks, taken from \citet{ps02}.  No clear trend is visible, although there is a slight trend for a higher ratio in older systems.  It may also be significant that the highest values of the line ratio are both for WTTS.

Fig.\ref{fig:m_gul} shows the line ratio plotted against mass accretion rate.  Mass accretion rate is difficult to measure, and only a small sample of objects have known accretion rates.  We have used the accretion rates derived by \citet{gul98}, derived by fitting accretion-shock models to observed spectra.  Other similarly-sized samples of mass accretion rates exist \citep{heg95,jkvl00}, but the systematic errors between the different methods used to derive the accretion rates mean that the data cannot be combined.  Again no strong trend is visible, but if anything the line ratio appears to increase as the objects evolve.

Fig.\ref{fig:ha} shows the line ratio plotted against H$\alpha$ equivalent width, as measured by \citet{ck79}.  The H$\alpha$ line arises due to accretion onto the central object, and thus decreasing H$\alpha$ emission is thought to be a good ``disc clock'' \citep{kitamura02}.  Again, no clear trend is visible, but we note again that the highest values tend to be for the WTTS.

In all cases the line ratio appears to increase, or at least not to decrease, as the objects evolve.  This appears to imply that the ionizing flux is not powered by mass accretion, as the fraction of the total flux emitted at short wavelengths, and consequently the observed He\,{\sc ii}/C\,{\sc iv} line ratio, would be expected to decrease dramatically in this case \citep{mjh03}.  Accretion shock models \citep{cg98} predict that the emission will decrease and peak at longer wavelengths as individual objects evolve, although an increased mass-to-radius ratio results in an intrinsically harder emitted spectrum.  However no correlation of any kind is observed between $M/R$ and the He\,{\sc ii}/C\,{\sc iv} line ratio and so we do not consider this to be a significant effect; both high and low line ratios are obtained from objects with both high and low $M/R$.  It is also significant that the highest values of the line ratio are obtained for the WTTS in the {\it IUE} sample, as these are expected to be disc-less objects and so disc accretion cannot be happening here.  Thus it seems likely that the production of ionizing photons around TTS is a stellar phenomenon, which we attribute to the chromosphere, and is independent of disc accretion.  As mentioned in Section \ref{sec:idea}, there is no evidence that magnetic activity in TTS is age-dependent, and so a large ionizing flux fraction should imply a strong absolute ionizing flux.  Thus it seems likely that TTS can produce an ionizing flux sufficient to power disc photoevaporation models: the ionizing flux is strong, and shows no significant decrease with time.


\section{Discussion}\label{sec:dis}
\subsection{Emission Measures}\label{sec:em_dis}
The use of the emission measure analysis is obviously crude, and contains many uncertainties.  The dominant uncertainty in the absolute values of $\Phi$ derived is the reddening uncertainty.  The manner in which the EMs are scaled also results in a form that is weakly sensitive to the choice of stellar parameters.  However this uncertainty is small compared to that introduced by reddening.

As regards the ``hardness'' of the spectrum, specifically the relative power radiated longward and shortward of the Lyman break, this is also not especially well constrained.  As mentioned in Section \ref{sec:idea} above, the ionizing/total power ratios derived are rather sensitive to the EM fits, and so the correlation noted in table \ref{tab:power_rat} is somewhat dubious.  We note however that the absolute values of $\Phi$ are not especially sensitive to the fit, with the fitting uncertainty remaining much less significant that that due to reddening.

However, we are able to reproduce both the observed line strengths, both in absolute and relative terms, to a good degree of accuracy.  Consequently we are satisfied that the EMs are consistent with the observed data, although we note that they are not always strongly constrained.  The absolute values of $\Phi$ derived in Section \ref{sec:EMs} are considered to be accurate only to around an order of magnitude, but they still provide a valuable new constraint.

\subsection{He\,{\sc ii} Line}\label{sec:heii_dis}
There are several obvious problems with the use of this line to infer the behaviour of the ionizing flux from TTS.  Firstly, and most importantly, the He\,{\sc ii} ionization energy is 54.4eV, and so recombination lines can only be used as an indicator of the flux of photons with energies greater than this (ie.~with wavelengths $<228$\AA).  Consequently this measure omits ionizing photons with wavelengths between 228\AA~and 912\AA, and photons in this wavelength range dominate the ionization of neutral hydrogen.  However it is difficult to envisage an emission process which could produce photons selectively in this wavelength region without producing higher-energy photons also, and so whilst the He\,{\sc ii} recombination line may not be ideal it should provide a reasonable indication of the flux ionizing neutral hydrogen.

Secondly there is the issue of the He\,{\sc ii} line itself, and whether it arises from radiative recombination (RR) or collisional excitation (CE).  This line has been observed in the solar corona, and has been explained both in terms of RR \citep{zirin75} and CE \citep{jordan75}.  \citet{linsky95,linsky98} have studied the He\,{\sc ii} 1640\AA~line in the chromosphere of Capella and found evidence for both mechanisms, but used high-resolution spectroscopy to show that the line profiles are characteristic of RR being the dominant factor.  More qualitatively, \citet{zirin75} found that his RR model works extremely well in active regions of the solar chromosphere, but less well in the quiet corona; given that T Tauri stars are characterised by their high levels of chromospheric activity it seems that RR will play a significant role.  Our synthetic spectra dramatically under-estimate the strength of the He\,{\sc ii} 1640\AA~line whilst reproducing the other lines and the continuum well, and whilst this lends further support to the RR theory we are unable to say with certainty whether the emission is CE or RR in origin.  However, it is worth bearing in mind that the densities and temperatures of plasma required to produce significant CE emission from He\,{\sc ii} are such that we would expect such a plasma to radiate significantly at wavelengths short of the Lyman break through free-free and bound-free transitions.  As a result, even if the emission is not RR, the presence of significant He\,{\sc ii} 1640\AA~emission should be a reliable indicator of a significant ionizing flux.  The 1640\AA~line is strong in almost all of the sources, both young and old, classical and weak-lined, and so it does seem that a significant ionizing flux is produced by all of these objects.

One should also note the fact that the ionization energy for He\,{\sc ii} is 4 times greater than that for H\,{\sc i}.  As a result the strength of the He\,{\sc ii} 1640\AA~line may be due to ionization by X-rays rather than softer ultraviolet photons.  Indeed, there is some evidence for a correlation between the He\,{\sc ii} 1640\AA~flux and X-ray power \citep{costa00}.  However this may merely indicate that both are good tracers of magnetic activity, rather than suggesting that the X-rays stimulate the He\,{\sc ii} 1640\AA~emission.  We are not able to make any distinction in this regard, but the conclusion that He\,{\sc ii} 1640\AA~line emission traces the production of energetic photons seems secure.

\subsection{C\,{\sc iv} Line}\label{sec:civ_dis}
There is also some uncertainty in the use of the C\,{\sc iv} 1550\AA~line as a normalising factor, as the relationship between C\,{\sc iv} flux and total radiated power is purely an empirical one (\cite{bc03}).  As mentioned above, the total radiated power is evaluated from an emission measure integral covering the UV and X-ray regimes.  They find that the X-ray contribution to this flux is ``less than 10\%'' of the total, but they do not state whether the ionizing or non-ionizing UV flux dominates the total.  However a previous paper \citep{costa00} indicates that the contributions are similar and that, if anything, the  non-ionizing UV flux is more significant.  Our emission measure analysis supports this, as the ionizing flux is not the dominant contribution to the total for any of the sources we consider (see Table \ref{tab:phi}).

It should also be noted that \citet{bc03} assume the C\,{\sc iv} emission to be optically thin.  If this is the case the relative intensities of the two components of the C\,{\sc iv} line should differ by a factor of 2 \citep{ardila02}.  We observe some evidence for opacity in 3 of the 9 objects observed with {\it STIS} (RY Tau, SU Aur and EZ Ori show ratios in the range 1.5-1.8).  We also noted possible evidence for opacity when comparing our synthetic spectra to the observed data (see Section \ref{sec:EMs}), although this may not be significant.  Therefore there is an unquantifiable degree of uncertainty here, as the resolution of the {\it IUE} data does not permit such analysis to be carried out.  We also note that a fraction of the C\,{\sc iv} line flux may be due to accretion, and so the slightly higher line ratios observed in the WTTS may not indicate a higher ionizing flux.  However the C\,{\sc iv} 1550\AA~line is one of the main coolants in T Tauri atmospheres/stellar chromospheres and so it is at least qualitatively reasonable to expect the relationship between C\,{\sc iv} flux and total radiated power to hold.  Moreover, \citet{bc03} found a strong correlation over nearly 2 orders of magnitude in power, and so we feel that the uncertainties due to the C\,{\sc iv} line are less of a concern than those due to the He\,{\sc ii} line discussed in Section \ref{sec:heii_dis}.

\subsection{Geometric issues}\label{sec:geom}
In addition to merely identifying the presence of a strong ionizing flux, there is also the problem of getting the photons to the disc.  As we have previously noted, the accretion columns are extremely optically thick to Lyman continuum photons \citep{columns}, and so it seems likely that any ionizing photons emitted by the chromosphere at low latitude will likely be absorbed by the columns.  In addition, any (bi-)polar outflow may also be optically thick to ionizing photons \citep{shang02}.  However in order to influence disc evolution it is only necessary to provide a strong ionizing flux to the disc at late times, and it is not clear if any outflow will persist to this stage of the evolution.  We also note that our diagnostic relies on a He\,{\sc ii} recombination line rather than a H\,{\sc i} recombination line, so we may see He\,{\sc ii} emission from regions where Lyman continuum emission is suppressed by absorption.

However there are also positives for the photoevaporation model.  Chromospheric emission at high latitude and/or from significantly above the photospheric surface will likely be unaffected by the accretion columns and, as mentioned above, outflows are much less significant at late times.  Also, whilst non-axisymmetric accretion columns \citep{mk98} will still absorb ionizing photons from the hotspot at their base they will permit much of the (roughly symmetric) chromospheric emission to escape unharmed.  Thus whilst the TTS geometry is not entirely favourable, it seems likely that most reasonable geometries will allow a significant fraction of the chromospheric ionizing flux to escape and influence the disc at late times.


\section{Summary}\label{sec:summary}
By analysing data from UV studies of TTS we have placed new constraints on the ionizing flux emanating from them.  An emission measure analysis using existing data has found rates of ionizing photons from the chromosphere in the range $\sim 10^{41}$--$10^{44}$ photons s$^{-1}$ for 5 CTTS with ages of around $10^6$yr.  It should be noted, however, that these values are subject to large observational uncertainties, in particular due to reddening.  We also note that our derived ionizing fluxes are intermediate between the solar value and the values typical of Herbig Ae/Be stars, and so seem qualitatively reasonable.

We have proposed that the He\,{\sc ii}/C\,{\sc iv} line ratio can be used as a reddening-independent diagnostic of the ionizing flux from TTS, essentially measuring the hardness of the spectrum.  We have analysed the behaviour of this line ratio in a larger sample of TTS observed by {\it IUE}, and find no evidence of any correlation with stellar or disc evolution.  Thus we conclude that the observed emission is likely due to the central object, presumably the stellar chromosphere, and is not tied to disc accretion.  This supports the hypothesis that the chromospheres of TTS can provide a strong ionizing flux.  Our method is somewhat crude and the results are by no means conclusive, but we have begun to place some constraints on a problem which was previously unbounded by observations.

These chromospheric Lyman continua appear to be sufficient to drive models of disc photoevaporation \citep{holl94,cc01,font04}.  However additional observations in the UV are clearly still needed to resolve this issue, as the current sample is small and the data are subject to large uncertainties.  In particular, studies of WTTS will be important in further constraining this problem, as all of our derived values of the ionizing flux are for CTTS.  The presence of a strong Lyman continuum around WTTS would provide further support for the hypothesis that the evolution of TTS discs is strongly affected by photoevaporation.

\section*{Acknowledgements}
We thank Bob Carswell for useful discussions.  We are grateful to David Brooks for providing his emission measure data in electronic form.  RDA acknowledges the support of a PPARC PhD studentship.  CJC gratefully acknowledges support from the Leverhulme Trust in the form of a Philip Leverhulme Prize.  RDA and CJC thank UC Santa Cruz for hospitality during the early stages of this work.  JEP thanks STScI for continued hospitality under their Visitors' Program.  {\sc chianti} is a collaborative project involving the NRL (USA), RAL (UK), and the Universities of Florence (Italy) and Cambridge (UK).  Based in part on observations made with the NASA/ESA Hubble Space Telescope, obtained from the data archive at the Space Telescope Science Institute. STScI is operated by the Association of Universities for Research in Astronomy, Inc. under NASA contract NAS 5-26555.
We thank an anonymous referee for several comments which improved the clarity of the paper.

\label{lastpage}

\end{document}